# Molecular Curvature Enables Tubular $\pi$-Stacked Pathways in Three-Dimensional Conductive Metal–Organic Frameworks

*Weishan Li[1], Jia-Xiang Zhang[1], Tongyang Zhao[1], Hao Chen[1,2], Tianyang Chen[3], Ze-Fan Yao[4] and Jin-Hu Dou[1]**

[1]National Key Laboratory of Advanced Micro and Nano Manufacture Technology; Key Laboratory of Polymer Chemistry and Physics of Ministry of Education, School of Materials Science and Engineering, Peking University, Beijing, 100871, China.

[2]Key Laboratory of Artificial Structure and Quantum Control, Ministry of Education, School of Physics and Astronomy, Shanghai Jiao Tong University, Shanghai, 200240, China.

[3] School of Science and Engineering, The Chinese University of Hong Kong (Shenzhen), Longgang, Shenzhen, Guangdong, 518172, China.

[4] Beijing National Laboratory for Molecular Sciences (BNLMS); Key Laboratory of Polymer Chemistry and Physics of Ministry of Education; Center of Soft Matter Science and Engineering, College of Chemistry and Molecular Engineering, Peking University, Beijing 100871, China.

* Corresponding author. Email: doujinhu@pku.edu.cn (J. D.)

**ABSTRACT:** Conductive metal–organic frameworks (MOFs) are commonly constructed from planar aromatic ligands, whose geometries favor conventional $\pi$-stacking arrangements. However, intrinsically twisted ligands offer an alternative geometric approach to directing aromatic packing and constructing unconventional framework architectures. Here, we employ intrinsically twisted, butterfly-shaped dibenzo[g,p]chrysene-2,3,6,7,10,11,14,15-octaol (DBC) ligands to construct a series of lanthanide MOFs. Single-crystal X-ray diffraction reveals an interpenetrated three-dimensional framework in which $\pi$–$\pi$ interactions between the aromatic fragments generate axially extended tubular $\pi$-stacked pathways, alongside one-dimensional channels rich in hydrogen-bonding sites. Lanthanide substitution preserves the overall tubular framework topology while modulating aromatic stacking distances and local pore environments. PrDBC single crystals exhibit mixed electron–proton transport, with electron and proton conductivities reaching $1.62 \times 10^{-3}$ and $4.37 \times 10^{-4}$ S cm$^{-1}$, respectively, at 338 K and 30% relative humidity. This work highlights molecular curvature as a structural design strategy for constructing three-dimensional conductive MOFs that integrate tubular aromatic stacks with tunable pore environments for mixed electron–proton transport.

## Introduction

The relationship between graphene and carbon nanotubes (CNTs) provides an intriguing example of how geometry influences the structural and electronic properties of $\pi$-conjugated systems[1-4]. Graphene features a planar, two-dimensional $\pi$-conjugated network, whereas CNTs can be viewed as graphene sheets rolled into tubular structures, introducing curvature while retaining an extended $\pi$-conjugated framework. The electronic structures and transport characteristics of CNTs depend on their tube diameter and chirality, defined by the chiral indices $(n,m)$[5, 6]. This relationship illustrates how geometric deformation can transform the spatial organization of a $\pi$-conjugated system and modulate its electronic properties, highlighting molecular curvature as a potentially powerful design parameter for constructing functional $\pi$-conjugated architectures.

Translating geometry-guided design from carbon-based $\pi$-conjugated materials to metal–organic frameworks (MOFs) offers opportunities to create structurally distinct arrangements of aromatic building blocks within crystalline coordination networks. In conductive MOFs, metal–ligand connectivity, intermolecular $\pi$–$\pi$ interactions, and framework topology collectively influence the arrangement of conjugated units and the resulting charge-transport properties. Three-dimensional frameworks offer greater flexibility in organizing aromatic building blocks in space and tailoring pore environments. Such structural versatility is particularly relevant to materials in which electronic transport must be integrated with other transport processes, such as proton conduction.

Despite these opportunities, many $\pi$-conjugated MOFs are constructed from planar or quasi-planar ligands, with their framework structures largely determined by metal–ligand connectivity and intermolecular packing, leading to 2D layered or quasi-2D structures [7, 8]. Although such building blocks have enabled the development of diverse conductive frameworks, exploiting the intrinsic curvature of twisted aromatic ligands to regulate intermolecular packing and provide additional structural freedom for constructing three-dimensional conductive MOFs remains relatively unexplored. Developing MOFs based on such ligands may thus provide new opportunities for controlling framework architectures and integrating complementary transport-related environments within a single framework.

Guided by this concept, we employ the intrinsically twisted $\pi$-conjugated ligand dibenzo[g,p]chrysene-2,3,6,7,10,11,14,15-octaol (DBC) to assemble a series of lanthanide MOFs (LnDBC) with trivalent lanthanide ions (Figure 1). Single-crystal structures of Ce-, Pr-, Nd-, and SmDBC reveal closely related three-dimensional interpenetrating frameworks in which the DBC ligands retain their characteristic nonplanar geometry. Aromatic fragments from neighboring interpenetrated components interact through $\pi$–$\pi$ stacking along the channel direction, forming axially extended tubular $\pi$-stacked pathways. The channels feature coordinated water molecules, disordered nitrate species, and uncoordinated hydroxyl groups, which collectively provide abundant hydrogen-bonding sites and create a hydrophilic pore environment[9-11]. Lanthanide substitution preserves the overall tubular framework topology while modifying aromatic stacking distances and local coordination environments. The coexistence of these structural features within a single crystalline framework provides a structural basis for mixed electron–proton transport. Motivated by these structural features, we further investigated the electron–proton transport properties of LnDBC. As a representative example, PrDBC single crystals exhibit relatively balanced electronic and protonic conduction, with conductivities of $1.62 \times 10^{-3}$ and $4.37 \times 10^{-4}$ S cm$^{-1}$, respectively, at 338 K and 30% relative humidity, highlighting the potential of LnDBC as mixed electron–proton conductors.

## Results and Discussion

### Synthesis and physicochemical characterization of LnDBC

A series of LnDBC MOFs was synthesized via a solvothermal method. In a typical procedure, hydrated $Ln(NO_3)_3$ was used as the metal precursor in excess relative to the intrinsically twisted DBC linker (Figure 2a). The reaction was carried out at 135 °C for 72 h in the presence of sodium acetate, which served as a modulator to facilitate ligand deprotonation and regulate crystal growth[12-14]. Detailed synthetic procedures are provided in the Supporting Information.

The obtained LnDBC crystals were initially characterized by optical microscopy (OM) and scanning electron microscopy (SEM). The samples generally exhibit irregular block-like morphologies, with dimensions of approximately 50–500 μm (Figures S1,2). PrDBC was selected as a representative sample for detailed physicochemical characterization. Fourier-transform infrared (FTIR) spectroscopy reveals a shift in the C–O stretching band from 1289 cm$^{-1}$ in pristine DBC to 1269 cm$^{-1}$ in PrDBC, consistent with ligand deprotonation and coordination to $Pr^{3+}$ centers (Figure S3)[15]. A nitrate-related vibrational band at 1384 cm$^{-1}$ further supports the presence of nitrate species in PrDBC. In addition, the broad O–H stretching feature is observed at approximately 3300 $cm^{-1}$ in pristine DBC and around 3500 $cm^{-1}$ in PrDBC, indicating changes in the local hydrogen-bonding environment involving hydroxyl groups and water molecules.

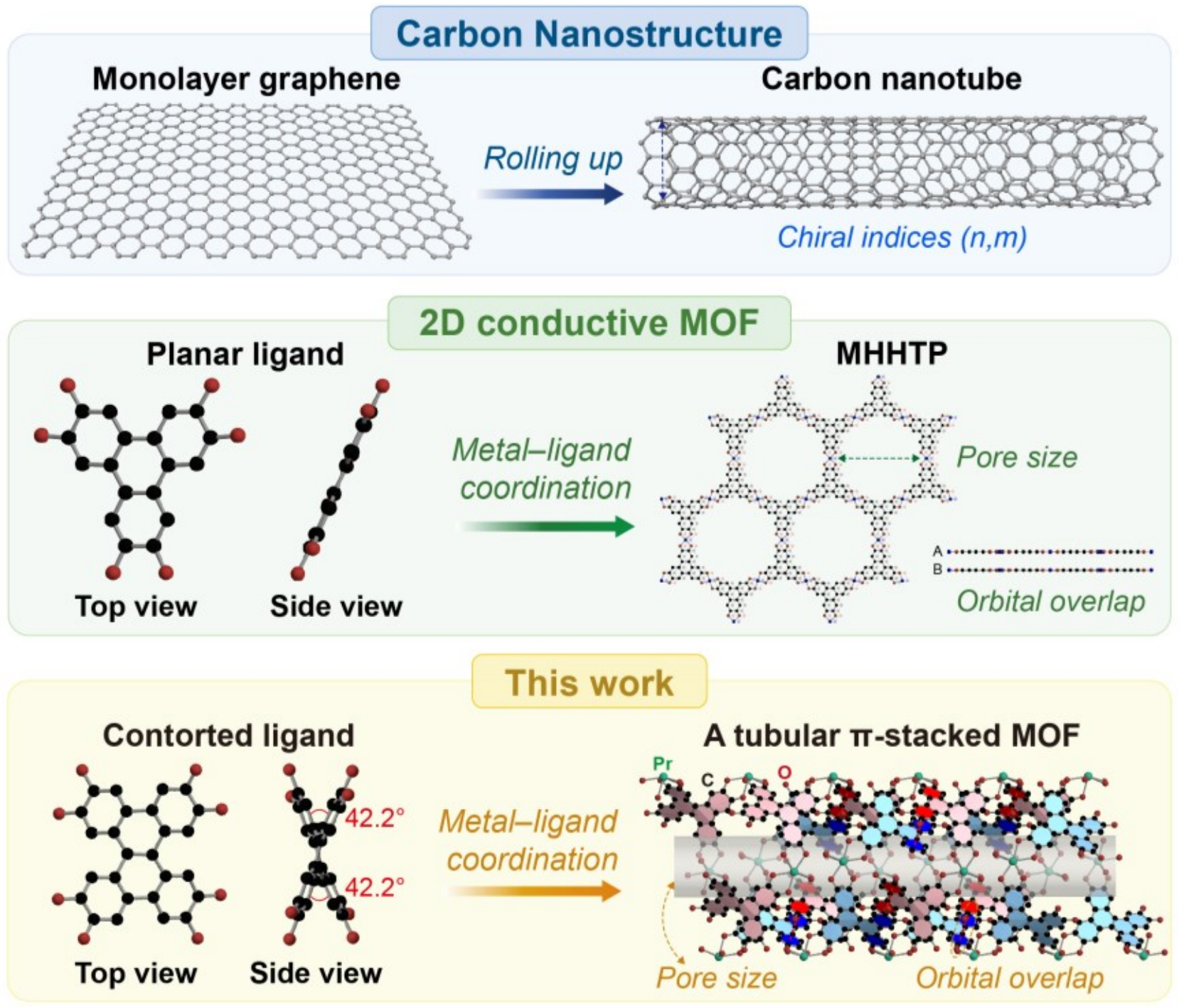

**Figure 1.** Ligand-geometry-guided construction of tubular $\pi$-stacked MOFs. Inspired by the geometric relationship between graphene and carbon nanotubes, this work introduces intrinsic ligand curvature into conductive MOF design. In contrast to the planar HHTP (2,3,6,7,10,11-Hexahydroxytriphenylene) ligands used to construct layered 2D MOFs, the twisted butterfly-shaped DBC ligand directs the assembly of interpenetrating lanthanide-based networks associated through $\pi$–$\pi$ interactions, generating a three-dimensional MOF architecture with axially extended tubular $\pi$-stacked pathways.

Energy-dispersive X-ray spectroscopy (EDS) confirms the presence of Pr, C, O, and N in PrDBC (Figure S4). X-ray photoelectron spectroscopy (XPS) was further employed to investigate the chemical states of the constituent elements (Figures S5 and S6). The Pr 3d spectrum exhibits spin–orbit-split features near 934 and 954 eV, corresponding to the Pr $3d_{5/2}$ and Pr $3d_{3/2}$ regions, respectively, consistent with the presence of $Pr^{3+}$ centers in the framework[16, 17]. The high-resolution C 1s spectrum contains components attributable to C–O and carbonyl-like bonding environments, with the latter potentially indicating partial oxidation of DBC-derived moieties[18]. A weak N 1s signal is also observed. Thermogravimetric analysis (TGA) reveals continuous mass loss without a well-defined plateau (Figure S7), which may reflect overlapping contributions from water loss, the release or decomposition of other labile species, and framework decomposition at elevated temperatures.

**Single Crystal Structure of LnDBC**

To elucidate the crystal structure of LnDBC, suitable single crystals were examined by single-crystal X-ray diffraction (SCXRD). As a representative example, PrDBC crystallizes in the monoclinic space group *C2/c* (No. 15) and adopts a three-dimensional interpenetrated framework constructed from twisted DBC ligands. The framework consists of two identical interpenetrating components related by a twofold symmetry operation. The experimental powder X-ray diffraction (PXRD) pattern of bulk PrDBC agrees well with that simulated from the single-crystal structure, supporting the structural consistency between the bulk sample and the single-crystal model (Figure S8).

In PrDBC, each $Pr^{3+}$ center is coordinated by six oxygen donor atoms from three DBC ligands, each contributing two donors, and one oxygen atom from an ordered water molecule. An additional water molecule and a bidentate nitrate anion occupy alternative positions in the coordination sphere, with refined occupancies of approximately 0.555 and 0.445, respectively (Figure 2b). Consequently, the $Pr^{3+}$ center exhibits a coordination number (CN) of eight or nine, depending on whether an additional water molecule or a bidentate nitrate anion is coordinated, respectively (Figure 2b). Each DBC ligand coordinates three $Pr^{3+}$ centers through three pairs of oxygen donors, while the remaining pair of hydroxyl groups is uncoordinated and exposed to the channel environment, providing potential hydrogen-bonding sites. The free DBC ligand exhibits a pronounced non-planar geometry, with a dihedral angle of 42.2° between the flanking benzene rings. This intrinsic distortion is largely retained in PrDBC, with the corresponding dihedral angles adjusted to 41.8° and 36.5°, and contributes to the formation of one-dimensional channels extending along the *c* axis (Figure 2b). The disordered water/nitrate coordination environment and uncoordinated hydroxyl groups provide polar sites within the channels, creating a hydrophilic pore environment. Nitrogen adsorption measurements at 77 K yield a Brunauer–Emmett–Teller (BET) surface area of 122.4 $m^2$ $g^{-1}$, supporting the pres ence of gas-accessible porosity (Figure 2c). The relatively modest surface area may be associated with partial pore occupation by framework-associated species and limited accessibility of the internal pore space to $N_2$.

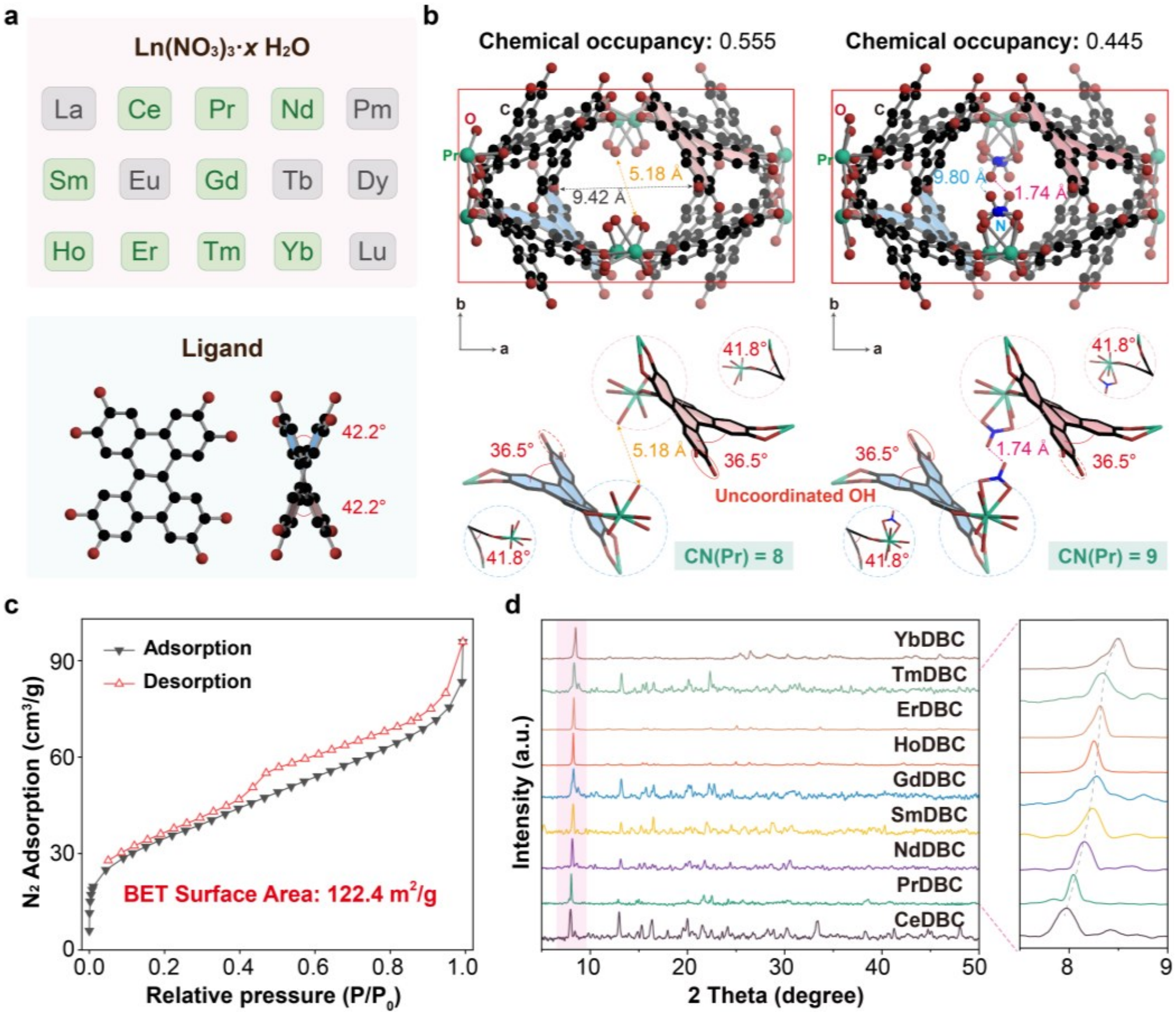

**Figure 2. Structural characterization of LnDBC frameworks.** (a) Lanthanide nitrate precursors and the intrinsically twisted DBC ligand used for LnDBC synthesis. Green indicates lanthanide metals for which LnDBC frameworks were obtained, while gray indicates other lanthanide metals. (b) Channel structure of PrDBC viewed along the *c* axis, showing $H_2O/NO_3^-$ coordination disorder at the Pr center, with refined occupancies of 0.555 and 0.445, respectively. (c) Nitrogen adsorption isotherm of PrDBC at 77 K. (d) PXRD patterns of LnDBC analogues (Ln = Ce, Pr, Nd, Sm, Gd, Ho, Er, Tm, and Yb). The enlarged region highlights the gradual shift of the main diffraction peak toward higher $2\theta$ values with decreasing lanthanide ionic radius, consistent with lanthanide contraction.

Beyond PrDBC, eight additional LnDBC MOFs (Ln = Ce, Nd, Sm, Gd, Ho, Er, Tm, and Yb) were synthesized by varying the lanthanide precursor. Their PXRD patterns exhibit similar diffraction features, suggesting closely related framework structures across the series. Notably, the main diffraction peaks gradually shift toward higher $2\theta$ values with decreasing lanthanide ionic radius, consistent with lattice contraction arising from lanthanide contraction (Figure 2d)[19]. SCXRD of Ce-, Nd-, and SmDBC further confirms that these compounds share the same three-dimensional interpenetrated framework topology as PrDBC.

In addition, a mixed-metal Ce/Pr/Sm-DBC framework was synthesized under the same solvothermal conditions using equimolar Ce, Pr, and Sm precursors. EDS elemental mapping reveals relatively uniform distributions of all three lanthanide elements within the examined crystal (Figure S9). Semiquantitative EDS analysis also indicates a higher proportion of Sm relative to Ce and Pr, potentially reflecting differences in metal incorporation during crystallization. These observations support the feasibility of incorporating multiple lanthanide ions into a common crystalline framework[20].

The three-dimensional architecture of LnDBC can be rationalized by the coordination flexibility of lanthanide ions in combination with the intrinsic nonplanarity of the DBC ligand. The highly localized and shielded 4f orbitals of $Ln^{3+}$ ions generally contribute little to directional metal–ligand bonding. The predominantly ionic character of Ln–O interactions permits diverse high-coordinate geometries, offering greater coordination flexibility than the square-planar coordination environments commonly encountered in 3d-metal-based conductive MOFs[21, 22]. Combined with the intrinsic molecular curvature of DBC, this coordination flexibility accommodates the twisted ligand geometry within a nonlayered three-dimensional framework.

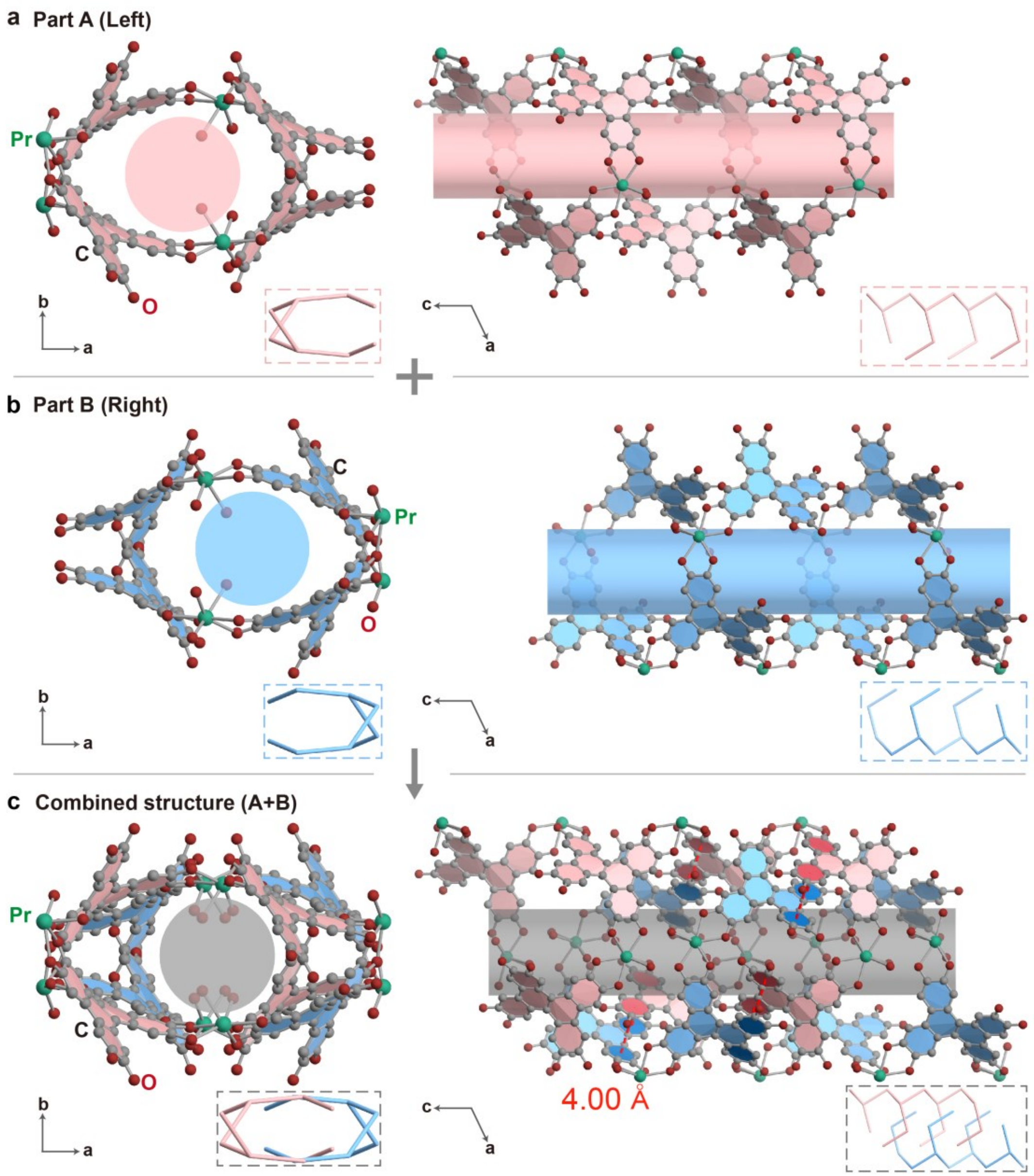

**Figure 3. Interpenetrating networks and tubular $\pi$-stacked pathways in PrDBC.** (a,b) Structural fragments of the two interpenetrating networks in PrDBC, highlighted in pink and blue, respectively. (c) Interpenetrated assembly of the two fragments, forming one-dimensional tubular channels along the *c* axis. The centroid–centroid distance between adjacent aromatic rings involved in offset $\pi$–$\pi$ stacking is approximately 4.00 Å.

Previous studies have reported three-dimensional conductive MOFs constructed from both planar[23] and intrinsically twisted aromatic ligands[8, 24]. However, translating molecular curvature into well-defined three-dimensional architectures that integrate extended $\pi$-stacking pathways with functional pore environments remains challenging. In this work, PrDBC adopts a distinct architecture comprising two interpenetrating networks (Figure 3a,b). Owing to the intrinsic curvature of DBC, aromatic fragments from neighboring networks form offset $\pi$-stacking contacts, with centroid–centroid distances of approximately 4.0 Å. These inter-network interactions organize the aromatic fragments into tubular $\pi$-stacked assemblies extending along the *c* axis, thereby defining one-dimensional channels within the framework (Figure 3c).

These two networks are related by a twofold rotation about an axis parallel to the *b* axis and collectively define the channel environment (Figure 3c). Coordinated water molecules, disordered nitrate species, and uncoordinated hydroxyl groups introduce local chemical heterogeneity within these channels. This architecture combines tubular aromatic stacks with polar, hydrogen-bond-rich channels, providing complementary structural motifs for electron and proton transport.

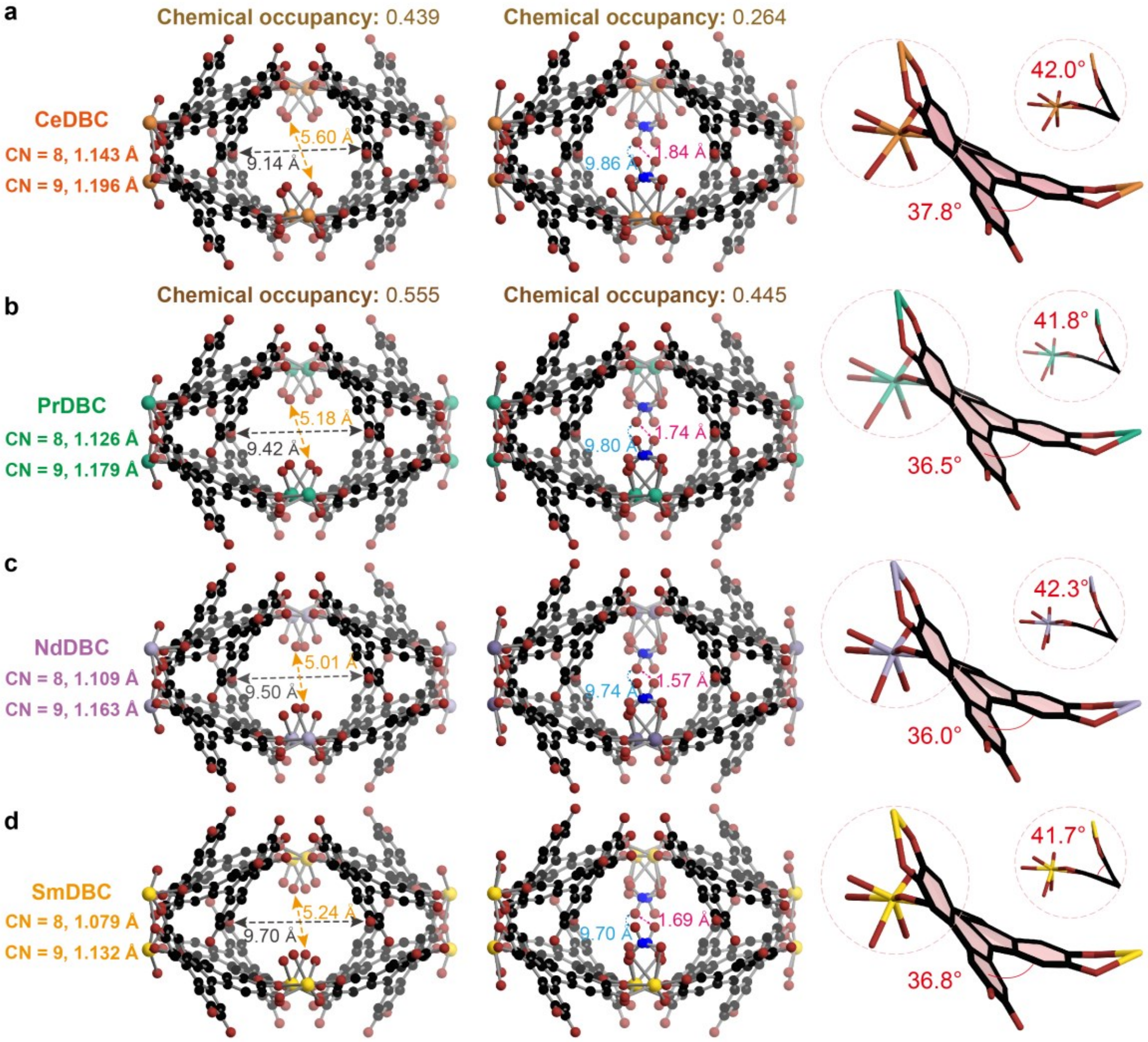

**Figure 4. Lanthanide-dependent channel geometry and local structural variation in LnDBC.** (a–d) Channel structures of Ce-, Pr-, Nd-, and SmDBC viewed along the *c* axis, highlighting the tubular $\pi$-stacked architectures, disordered water/nitrate species and their refined occupancies, coordination environments, and selected channel-related local distances. The DBC dihedral angles illustrate the preservation of ligand curvature across the series.

To examine the electrostatic consequences of the $H_2O/NO_3^-$ coordination disorder, two idealized local coordination models were constructed from finite PrDBC fragments, with the disordered coordination sites occupied by either $H_2O$ or $NO_3^-$. Their electrostatic potential (ESP) distributions were calculated using the xTB method (Figure S10). The two models exhibit distinct ESP distributions around the Pr coordination sites and within the channel interior, whereas the ESP distribution over the DBC backbone remains qualitatively similar. These model calculations suggest that the crystallographically observed $H_2O/NO_3^-$ disorder may give rise to local electrostatic heterogeneity within the tubular channels. Such heterogeneity could influence the organization of confined water molecules and the associated hydrogen-bonding interactions relevant to proton transport.

To further assess the preservation of the tubular architecture upon lanthanide substitution, we compared the single-crystal structures of Ce-, Pr-, Nd-, and SmDBC (Figure 4, Figures S11–S14 and Tables S1–S4). All four compounds share the same three-dimensional interpenetrated framework topology and retain the curvature-defined tubular architecture. The DBC ligands maintain their characteristic curved geometry, with representative dihedral angles of approximately 37° and 42°. While Pr-, Nd-, and SmDBC exhibit closely related structures, CeDBC displays more pronounced local disorder, including positional disorder of the uncoordinated aromatic segments of DBC and a more complex pattern of $H_2O/NO_3^-$ occupancy and local coordination within the channels (Figures S15 and S16).

As the lanthanide ionic radius decreases from $Ce^{3+}$ to $Sm^{3+}$, the Ln–O distances show an overall shortening trend, accompanied by a gradual decrease in unit-cell volume (Figure S17). However, LnDBC does not undergo simple isotropic contraction; instead, the framework accommodates lanthanide-dependent structural changes through coupled adjustments in lattice parameters, the relative arrangement of the interpenetrating networks, and local coordination environments. Across the series, the centroid–centroid distance between adjacent, offset-stacked DBC aromatic fragments decreases from 4.04 to 3.93 Å, indicating closer aromatic contacts within the tubular $\pi$-stacked pathways. Such changes in aromatic packing may modulate intermolecular electronic coupling and thereby affect electronic transport. Meanwhile, short-range contacts involving coordinated water molecules, nitrate species, and uncoordinated hydroxyl groups also vary with the lanthanide center, indicating that metal-ion substitution can modulate the local hydrogen-bonding environment within the channels.

To further examine the structural adaptability of the tubular framework, we modified the crystallization conditions by replacing $Pr(NO_3)_3$ with $PrCl_3$ or lowering the reaction temperature to 105 °C. Under these conditions, the overall framework topology was preserved, while the DBC ligands exhibited distinct local conformations (Figures S18–S20). Across these structures, one characteristic dihedral angle remained close to 39°, whereas the other varied from 35.6° to 45.3°, indicating localized torsional adjustments. These observations demonstrate that the intrinsically curved DBC ligand can undergo local conformational adjustments without disrupting the tubular $\pi$-stacked architecture.

**Electron and Proton Conductivity**

The tubular aromatic stacking and hydrogen-bond-rich channels in LnDBC provide complementary structural motifs for electron and proton transport, motivating us to investigate mixed electron–proton conduction. The tubular $\pi$-stacked pathways generated by the curved DBC ligands provide extended aromatic contacts along the crystallographic $c$ axis, while the adjacent hydrogen-bond-rich channels containing coordinated water molecules, nitrate species, and uncoordinated hydroxyl groups provide an environment potentially conducive to proton migration.

We first evaluated the optical absorption properties of the LnDBC series by ultraviolet−visible spectroscopy (UV−vis). Tauc analysis of PrDBC gives an optical gap of approximately 0.855 eV, while the other LnDBC analogues exhibit similar values near 0.850 eV(Figures S24 and S25). These comparable optical gaps are consistent with their closely related framework connectivity and similar ligand coordination environments. Considering the localized nature of lanthanide 4f orbitals and their limited overlap with ligand $\pi$ orbitals, the low-energy absorption may involve transitions associated with DBC-derived $\pi$-electronic states.

We next evaluated the electronic conductivity of LnDBC powder pellets using the van der Pauw method. As the lanthanide ionic radius decreases, the pellet conductivity gradually increases from $3.37 \times 10^{-7}$ to $1.03 \times 10^{-5}$ S cm$^{-1}$ (Figures 5a, S26 and S27). This conductivity variation correlates with the lanthanide-dependent structural evolution of the tubular $\pi$-stacked framework. Specifically, comparison of the Ce-, Pr-, Nd-, and SmDBC structures shows that the centroid–centroid distance between adjacent DBC aromatic fragments decreases from 4.04 to 3.93 Å, indicating closer aromatic contacts and modified $\pi$-stacking geometry within the tubular pathways. Such changes in aromatic packing may influence intermolecular electronic interactions and contribute to the observed conductivity enhancement across the LnDBC series. Nevertheless, because powder-pellet measurements are affected by grain boundaries, interparticle contacts, pellet density, and possible preferred orientation, these values represent effective pellet conductivities rather than intrinsic single-crystal conductivities.

To probe directional electronic transport in the tubular framework, relatively large PrDBC single crystals were selected for electrical measurements. The crystals were oriented along the crystallographic $c$ axis, corresponding to the direction of the tubular $\pi$-stacked pathways, and their conductivity was measured using a two-probe configuration. PrDBC single crystals exhibit an apparent room-temperature conductivity on the order of $10^{-5}$ S cm$^{-1}$ along the $c$ direction (Figure S28). Other LnDBC single crystals show conductivities within the same order of magnitude, indicating comparable electronic transport behavior across the series. The absence of a clear monotonic trend in the single-crystal conductivity data may reflect uncertainties in crystal dimensions and alignment, differences in defect populations, and variations in electrode contact resistance.

Taking advantage of the relatively large and robust PrDBC single crystals, we further investigated mixed electron–proton conduction at the single-crystal level. The larger crystal dimensions facilitated stable electrode contact and helped minimize mechanical damage during device fabrication[25], enabling more reliable transport measurements. Linear sweep voltammetry (LSV) and electrochemical impedance spectroscopy (EIS) were combined to analyze charge transport under controlled humidity and temperature conditions.

The EIS spectra of PrDBC show two distinguishable semicircles, consistent with two distinct relaxation processes[26, 27]. By combining LSV measurements with equivalent-circuit fitting of the EIS spectra, the electron and proton transport contributions can be separated within the adopted equivalent-circuit model. The electronic resistance determined from LSV is consistent with that extracted from EIS fitting, supporting the electronic resistance assignment in the equivalent-circuit model.

Humidity-dependent measurements reveal the coexistence of electron and proton conduction in PrDBC. At 25 °C and 30% relative humidity, the electron and proton conductivities are $2.44 \times 10^{-5}$ and $4.72 \times 10^{-5}$ S cm$^{-1}$, respectively. Upon increasing the relative humidity to 90%, these values increase to $2.91 \times 10^{-5}$ and $2.21 \times 10^{-4}$ S cm$^{-1}$, respectively (Figures 5b,c). The electronic conductivity shows only a moderate increase, which may arise from the dielectric screening effect of adsorbed water molecules and the resulting modulation of charge-carrier transport within the framework[28]. In contrast, the proton conductivity increases more substantially, suggesting that pore water facilitates proton transport, potentially by improving hydrogen-bond connectivity within the channels. Notably, the electron and proton conductivities are of comparable orders of magnitude, demonstrating relatively balanced mixed electron–proton conduction within a single-crystalline framework.

Temperature-dependent measurements of PrDBC at 30% relative humidity reveal thermally activated transport behavior, with electron and proton conductivities increasing from $5.59 \times 10^{-4}$ and $8.41 \times 10^{-5}$ S cm$^{-1}$ at 25 °C to $1.62 \times 10^{-3}$ and $4.37 \times 10^{-4}$ S cm$^{-1}$ at 65 °C, respectively (Figures 5d,e). Arrhenius analysis gives activation energies of 0.209 eV for electron conduction and 0.370 eV for proton conduction (Figure 5f). The proton-transport activation energy is consistent with a hydrogen-bond-assisted, Grotthuss-like proton-hopping process[9], in which protons migrate through hydrogen-bonding networks formed by pore water molecules, uncoordinated hydroxyl groups, and other proton-accepting sites (Figure 5g), rather than relying primarily on a vehicle-type diffusion mechanism.

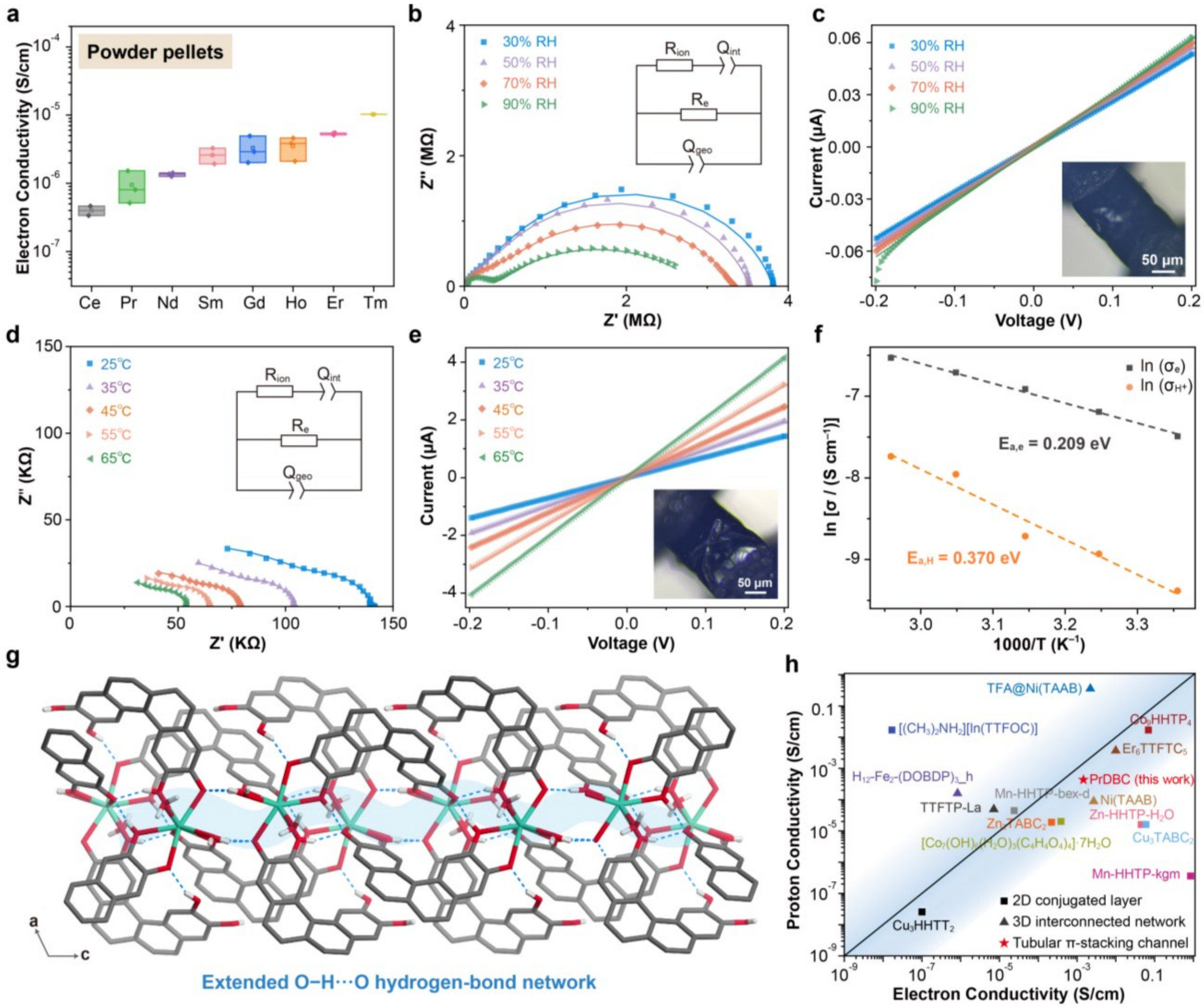

**Figure 5.** Mixed electron-proton transport in LnDBC. (a) Electron conductivities of LnDBC powder pellets. (b) EIS spectra and (c) LSV curves of PrDBC measured at 25 °C under different relative humidity conditions. Solid lines represent fitted curves. (d) EIS spectra and (e) LSV curves of PrDBC measured at 30% RH under different temperatures. (f) Arrhenius plots of temperature-dependent electron and proton conductivities. (g) SCXRD-resolved hydrogen-bond network in the channels of PrDBC, viewed along the crystallographic *b* axis. (h) Benchmarking the proton-electron conductivity of PrDBC with previously reported mixed-conducting MOFs under high-humidity conditions. Literature values were collected from $[(CH_3)_2NH_2][In(TTFOC)]$[29], $Cu_3HHTT_2$[30], $H_{12}$-$Fe_2$-$(DOBDP)_3$_h[31], TTFTP-La[32], Mn-HHTP-bex-d[33], Mn-HHTP-kgm[33], $Zn_3TABC_2$[34], $Cu_3TABC_2$[34], $[Co_7(OH)_6(H_2O)_3(C_4H_4O_4)_4]\cdot 7H_2O$[35], TFA@Ni(TAAB)[36], Zn-HHTP-$H_2O$[37], and $Er_6TTFTC_5$[38].

Together with the EIS-resolved electronic and protonic contributions, these results show that PrDBC integrates tubular aromatic-stacking pathways for electronic transport and hydrogen-bond-rich channels for proton transport within a single crystalline framework. Compared with previously reported MOF-based mixed conductors, PrDBC achieves a favorable balance between electron and proton conductivity, with both contributions reaching relatively high levels among crystalline framework materials under comparable conditions (Figure 5h). This balanced electron–proton conductivity is particularly important for processes requiring coupled transport of both carriers, where the less conductive component may limit overall transport performance[39, 40]. These results highlight lanthanide-based MOFs as a promising platform for designing structurally defined mixed electron–proton conductors.

## Conclusion

In summary, this work demonstrates molecular curvature as a structural design principle for constructing non-layered three-dimensional conductive MOFs. The intrinsically curved DBC ligand translates molecular geometry into framework-level interpenetration and tubular $\pi$-stacked architectures. Across the LnDBC series, the framework connectivity is largely preserved, while lanthanide substitution modulates lattice parameters, aromatic stacking, and the channel microenvironment. These findings demonstrate that ligand geometry and metal coordination flexibility can be combined to construct three-dimensional architectures beyond the conventional planar-ligand and layered-stacking paradigm. The tubular $\pi$-stacked aromatic pathways and hydrophilic, hydrogen-bond-rich channels provide complementary structural motifs for mixed electron–proton conduction. In PrDBC single crystals, combined LSV and EIS measurements resolve electron and proton contributions within the same crystalline framework, with conductivities reaching $1.62 \times 10^{-3}$ and $4.37 \times 10^{-4}$ S cm$^{-1}$, respectively, at 338 K and 30% relative humidity. The comparable magnitudes of electron and proton conductivity demonstrate relatively balanced mixed conduction within a single crystalline framework. More broadly, LnDBC provides a single-crystalline model platform for correlating molecular conformation, framework interpenetration, and pore microenvironment with mixed electron–proton transport. This design concept will guide the development of crystalline framework materials for proton-coupled electronics, electrochemical interfaces, and other applications requiring integrated ionic and electronic conduction.

## ASSOCIATED CONTENT

### Supporting Information

The following files are available free of charge.

Complete experimental procedures and other supporting data. (PDF)

The crystallographic data have been deposited at the Cambridge Crystallographic Data Centre (CCDC) with accession numbers CCDC 2583712-2583715 and 2587174-2587176 (CIF).

## AUTHOR INFORMATION

### Corresponding Author

*Jin-Hu Dou
doujinhu@pku.edu.cn (J. D.)

### Author Contributions

### Funding Sources

This work was financially supported by the National Key R&D Program of China (Grant No. 2023YFE0206400), National Natural Science Foundation of China (Grant No. 22171185, 22575005), and the Clinical Medicine Plus X-Young Scholars Project at Peking University backed by the Fundamental Research Funds for the Central Universities.

### Notes

The authors declare no competing interests.

### ACKNOWLEDGMENT

This work was financially supported by the National Key R&D Program of China (Grant No. 2023YFE0206400), National Natural Science Foundation of China (Grant No. 22171185, 22575005), and the Clinical Medicine Plus X-Young Scholars Project at Peking University backed by the Fundamental Research Funds for the Central Universities. The support of the beamline staff at BL02U2, BL10U1, BL14B1 and BL17B of SSRF is acknowledged for their assistance with data acquisition and analysis. We gratefully acknowledge Jie Su and Anqi He of the Analytical Instrumentation Center at Peking University for their assistance with single-crystal X-ray crystallographic analysis, and Prof. Qingqing Ji and Qingsheng Zeng of ShanghaiTech University for their valuable support. The authors acknowledge the High-Performance Computing Platform of Peking University for supporting computational work.

## SYNOPSIS TOC

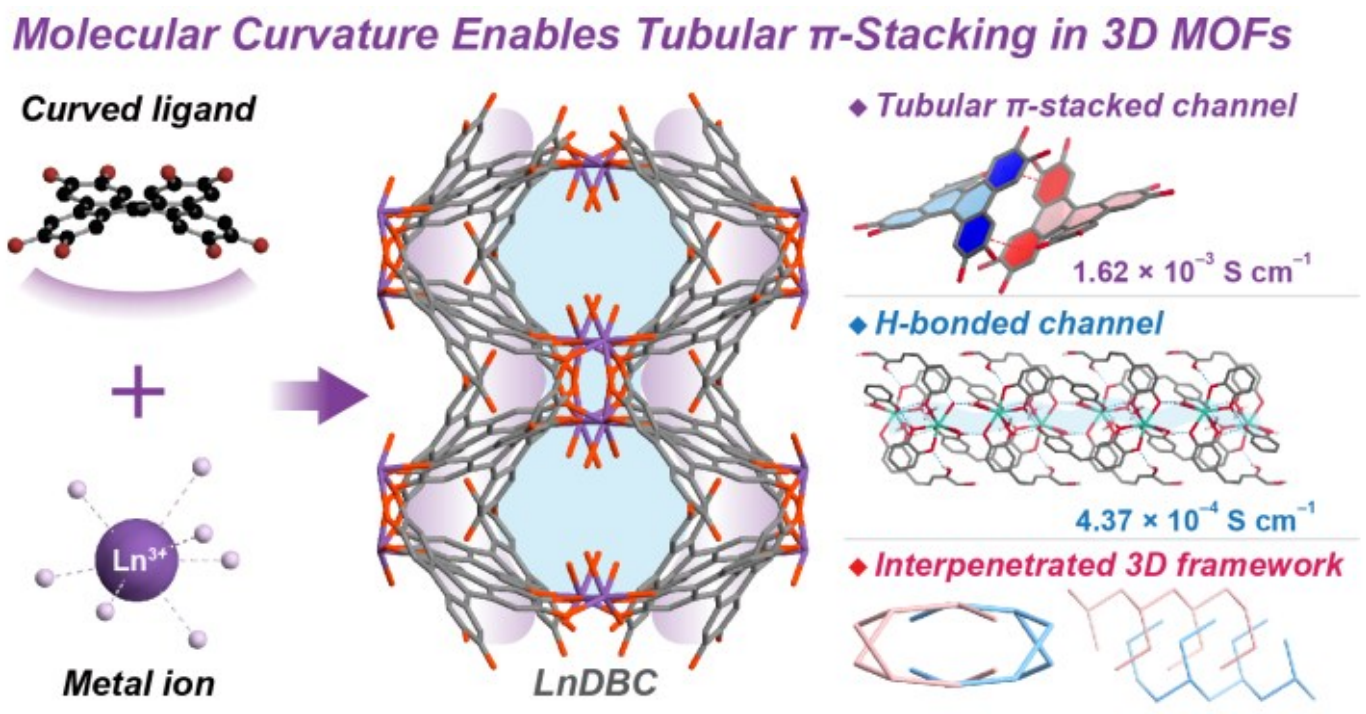